\documentclass[12pt]{article}
\usepackage{relsize}
\usepackage{wrapfig}
\usepackage{lscape}
\usepackage{longtable}
\usepackage{caption}
\usepackage{svg}
\usepackage{float}
\usepackage[utf8]{inputenc}
\usepackage{fullpage}
\usepackage{indentfirst}
\usepackage{amsmath}
\usepackage{graphicx}
\graphicspath{{./images/}}
\usepackage[backend=biber, sorting=none]{biblatex} 
\makeatletter

\usepackage{hyperref}
\usepackage{xr-hyper}
\title{A web-based user interface for Fam3PRO, \\ a multi-gene, multi-cancer risk prediction model \\ for families with cancer history}

\begin{document}

\maketitle


\noindent Xueying Chen\textsuperscript{1,2},
Jianfeng Ke\textsuperscript{3}, Lauren Flynn\textsuperscript{4}, Giovanni Parmigiani\textsuperscript{1,2}, Danielle Braun\textsuperscript{1,2}

\begin{enumerate}
    \item Dana-Farber Cancer Institute, 450 Brookline Ave, Boston, MA 02215, USA
    \item Harvard T.H. Chan School of Public Health, 677 Huntington Ave, Boston, MA 02115, USA
    \item Department of Biological Sciences, University of Massachusetts Lowell, Lowell, MA 01854, USA
    \item Division of Pulmonary Medicine, Boston Children's Hospital, Boston, MA 02115, USA
\end{enumerate}

Danielle Braun, PhD; email: dbraun@mail.harvard.edu

Running head: Fam3PRO UI for Multi-Gene and Multi-Cancer Risk Prediction

\section*{Abstract}

\subsection*{Purpose}
Hereditary cancer risk is key to guiding screening and prevention strategies. Cancer risks can vary by individual due to the presence or absence of high- and moderate-risk pathogenic variants (PV) in cancer-associated genes, in addition to sex, age, and other risk factors. We previously developed Fam3PRO, a flexible multi-gene, multi-cancer Mendelian risk prediction model that estimates a patient's risk of carrying a PV in hereditary cancer genes and their future risk of developing several types of cancer. The Fam3PRO R package includes 22 genes with 18 associated cancers, allowing users to build customized sub-models from any gene–cancer set. However, the current R package lacks a user interface (UI), limiting its practical use in clinical settings. Therefore, we aim to develop a web-based UI for broader use of the Fam3PRO functionalities.  

\subsection*{Methods}
The Fam3PRO UI (F3PI), built with R Shiny, collects and formats inputs including family health history, genetic test results, and other risk factors. Pedigree data are interactively visualized and modified via pedigreejs, while the backend Fam3PRO model takes all the inputs to generate carrier probabilities and future cancer risks, presented through an interactive UI.

\subsection*{Results}
F3PI streamlines the collection of patient and family history data, which is analyzed by the Fam3PRO models to provide personalized cancer risks for each proband across 18 cancers, as well as probabilities that a proband has a PV in up to 22 hereditary cancer genes. These results are returned to the user, within one minute on average and are available in both interactive and downloadable formats. 

\subsection*{Conclusion}
We have developed F3PI, an easy-to-use, interactive web application that makes cancer and genetic risk information more accessible to providers and their patients.

\section*{Introduction}

Hereditary cancer syndromes (HCSs) arise from mutations that confer an elevated susceptibility to cancer, and are the most frequent among Mendelian genetic diseases \cite{Garutti2023-ea, Imyanitov2023-ca}. For example, Lynch syndrome and hereditary breast and ovarian cancer (HBOC) are two of the HCSs with high morbidity. Determining how likely a proband carries pathogenic or likely pathogenic variants in cancer susceptibility genes could have substantial implications in terms of testing procedures, tailored prevention strategies, and treatment plans for different patients.

Given these clinical implications, a variety of estimation tools with user interfaces (UIs) have been developed to assess future cancer risks, utilizing numerous underlying statistical models, and designed to support both patients and healthcare professionals. Some of the tools predict cancer risks related to germline mutations. For example, ASK2ME, an online cancer risk calculator, uses models to assess cancer risk associated with pathogenic variants in various cancer susceptibility genes in multiple cancers using a robust knowledge base \cite{Braun2018-mi}. Similarly, MyLynch is a patient-facing tool designed for individuals with Lynch syndrome. It assesses cancer risk based on models that estimate age-specific conditional penetrance, focusing on genes specifically associated with the condition \cite{Knapp2023-me}. Both tools target post-genetic testing risk counseling for individuals with PVs and do not take into account family health history (FHx). However, FHx is important for risk counseling when genetic testing or germline mutation information is unavailable, or when germline genetic testing results indicate variants of unknown significance (VUS) or potentially for those with benign or no variants. 

While some tools focus solely on individuals with known PVs and do not consider FHx, others have been developed to estimate cancer risk based on FHx, particularly when genetic testing results are unavailable or inconclusive. For example, the Breast Cancer Risk Assessment Tool \cite{National-Cancer-InstituteUnknown-wk}, uses the Gail model, which does not consider PVs information, but computes the overall risk of developing invasive breast cancer, using FHx (the number of first-degree relative to breast cancer), as well as hormonal and reproductive risk factors \cite{Gail1989-pt}. Other tools that calculate breast cancer risks use the IBIS/Tyrer-Cuzick model and are free patient-facing online user interfaces, with licensed applications for clinical use \cite{IBIS-IkonopediaUnknown-vr, MagViewUnknown-hr}. The IBIS/Tyrer-Cuzick model itself estimates a 10-year and lifetime risk of cancer by incorporating extended FHx (beyond first-degree relatives), personal history of breast or ovarian cancer, and results of the genetic tests of BRCA1 / BRCA2 \cite{Tyrer2004-wc, IBIS-IkonopediaUnknown-vr}. 

A wide range of FHx-based tools that provide cancer risk estimates varying in function, scope, and level of detail have been systematically reviewed and analyzed \cite{Cleophat2018-pv, Welch2018-ds, Mirosevic2022-io}. For example, the MeTree tool includes multiple models (BRCAPRO, MMRpro, and PREMM) to estimate cancer risks and carrier probabilities for varius HCSs including breast, ovarian, and colon cancer  \cite{Orlando2013-kk, Guan2023-ii, Kastrinos2017-he, Chen2006-ng}. Other commercially used risk assessment solutions, such as MyRisk and CancerNext, run genetic lab tests and risk modeling for multiple cancer-associated genes across various cancer types, including FHx. Although they feature comprehensive UIs for providers and patients and offer \cite{Genetics2022-fs, GeneticsUnknown-xy} a variety of genes and cancer types to test, these tools are not public-facing and are accessible only in clinical or institutional settings.

Despite the availability of these tools, the lack of a standardized procedure for collecting and displaying FHx poses challenges to the accuracy and consistency of FHx-based risk assessment. Patients often have limited knowledge of their family history, and the actionable information derived from it can vary between clinicians according to their experience and interpretation \cite{Cleophat2018-pv, Orlando2013-kk}. In addition, constructing, modifying and visualizing a proband's family history, typically represented as a pedigree, can be challenging. An interactive editing tool, pedigreejs \cite{Carver2017-uo}, effectively addresses these issues. Pedigreejs is a lightweight JavaScript-based pedigree editor that generates interactive SVG images directly in the browser, allowing users to input and edit FHx data with a variety of features. Using pedigreejs for the collection and visualization of FHx in current cancer risk estimation tools, patients, clinicians, and genetic counselors can participate in an interactive and user-friendly process to build and update family history information. 

One tool that has successfully used pedigreejs is Canrisk, which is physician-facing, has a well-established user interface, and applies BOADICEA as the prediction model \cite{Lee2019-au}. BOADICEA integrates both germline genetic information, such as polygenic risk scores (PRS), and other factors (including FHx and mammographic density) in a comprehensive way to estimate carrier probabilities for breast and ovarian cancer \cite{Lee2019-au, Carver2021-rb}. Although Canrisk has nicely integrated FHx collection, risk factors, and a prediction model, it has limitations in the types of cancer and HCSs it covers. This highlights the need for an accessible and comprehensive tool capable of estimating cancer risks across a broad range of cancer types and associated genes. To address this gap and streamline risk estimation for diverse gene sets and cancers, we first developed Fam3PRO (formerly PanelPRO) a multi-gene, multi-cancer risk model designed for families with a history of cancer or relevant gene mutations \cite{Lee2021-xx}. Fam3PRO collects FHx and integrates it with demographic, clinical and genetic risk factors to estimate future cancer risks, as well as posterior carrier probabilities of HCS genes compared to an averaged person \cite{Lee2021-xx}. However, using Fam3PRO requires knowledge of the programming language R and currently there is no Fam3PRO user interface available, which limits its usability in daily clinical operations and restricts its use by patients, clinicians, and genetic counselors. 

Therefore, to broaden Fam3PRO's usability, we utilized the programming tools R Shiny, pedigreejs, and Fam3PRO to create a web application called the Fam3PRO Interface (F3PI), which is public-facing, and addresses the aforementioned issues. F3PI enables clinicians to build a family tree with detailed FHx in a standardized pedigree format, which can be easily used as input for Fam3PRO models and easily stored for future use. After building the pedigree, F3PI runs Fam3PRO and displays the model results in approximately one minute, with multiple formats of the output available for download. 

\section*{Methods}

\subsection*{Software Stack}

\begin{figure}[H]
\centering
\includegraphics[width=0.8\linewidth]{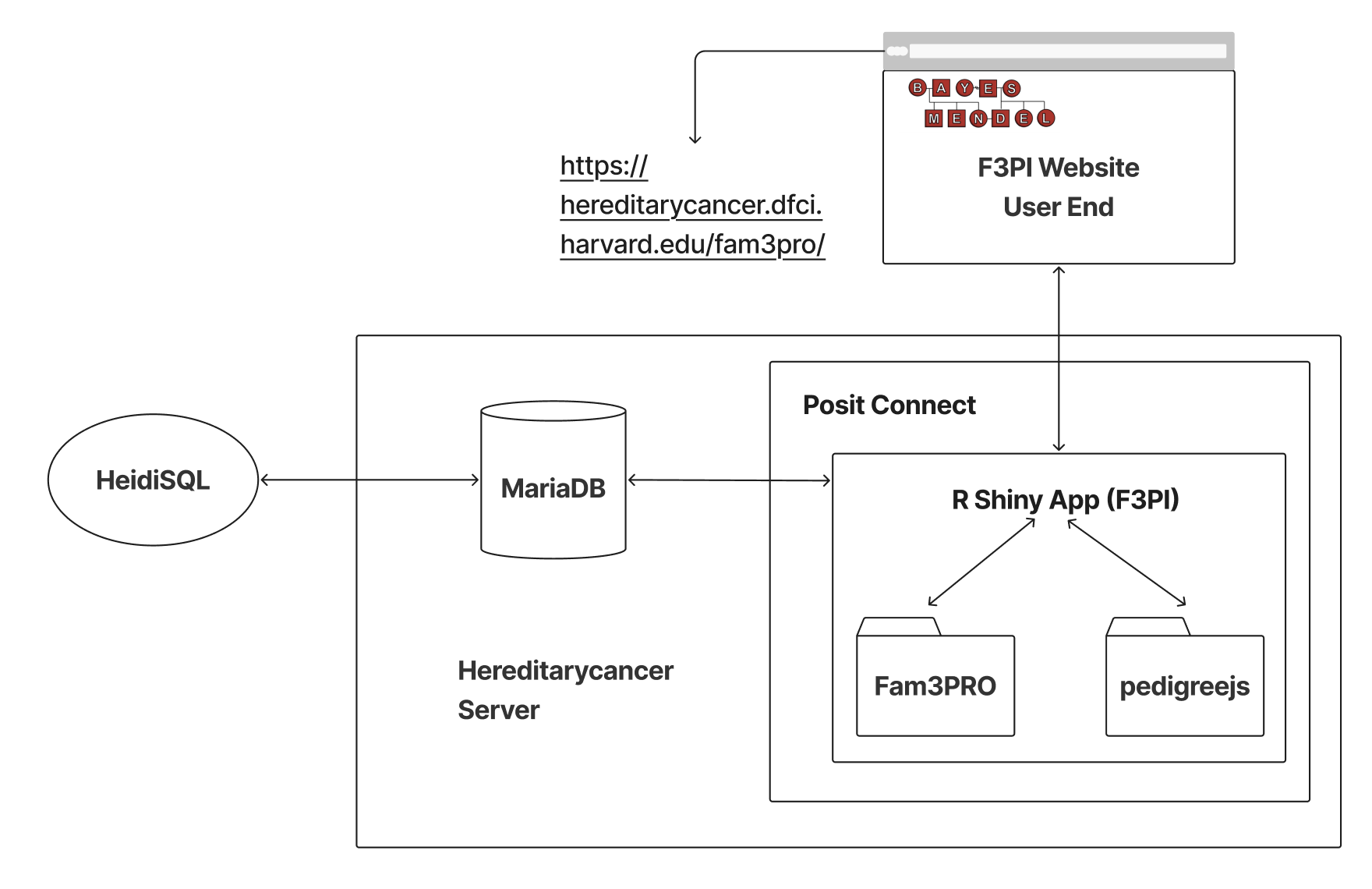}
\caption{Software architecture of the Fam3PRO UI (F3PI). The MariaDB database and the F3PI app, deployed on the Posit Connect platform, are both hosted on the hereditarycancer server at the Dana-Farber Cancer Institute. The Fam3PRO R package is cached on the server upon deployment. Users can interact with the R Shiny app and other components through the F3PI website.}
\label{fig1}
\end{figure}

Since pedigrees rely on human input and require extensive details about many family members, we sought an interface that was interactive, fast, and accurate. F3PI was programmed in R (version 4.4.1) using shiny (version 1.9.1) \cite{Campbell2024-hm}, an R package that enables the creation of interactive web applications and the publication on publicly accessible websites. The R Shiny application is deployed and hosted on a Posit(RStudio) Connect platform \cite{UnknownUnknown-ff}, and hosted on a Rocky Linux 8.10 server, to which we refer as the {\it hereditarycancer server}, at the Dana-Farber Cancer Institute. Users interact with it through a publicly accessible website URL (Figure ~\ref{fig1}). 

To construct family pedigrees, the application integrates pedigreejs, an open-source JavaScript-based software, into the Shiny-based web application. Pedigreejs handles the pedigree structure and displays an interactive family tree, allowing users to easily edit it, for example by adding or removing family members \cite{Carver2017-uo}. 

To store pedigrees and relevant information for each relative in the pedigree, a MariaDB database hosted on the hereditarycancer server as the application is linked to the F3PI. MariaDB is managed through HeidiSQL (version 12.10), which is an open-source graphical interface for database connection and administration \cite{Becker2025-yi}. R shiny communicates with MariaDB through the R package DBI (version 1.2.3) and RMariaDB (version 1.3.2) \cite{R-Special-Interest-Group-on-Databases-R-SIG-DB-2024-fl, Muller2024-dg}, which allows it to access the relevant stored information needed to run the Fam3PRO model. 

With user input from the website interface for each proband, the Shiny-based script interacts with pedigreejs to build pedigrees, save data to the MariaDB database, and call the Fam3PRO R package to run the risk models in the back-end, generating the final interactive visualizations of the model outputs in the interface (Figure ~\ref{fig1}). 

\subsection*{Pedigree Builder}

The pedigree builder is capable of building a fully linked family tree structure, which makes adding and removing relatives simple with minimal user entries. In addition, detailed data collection for each relative in a family is integrated into the pedigree builder to generate model inputs of cancer-related risk factors such as demographics, cancer history, prophylactic surgical history, tumor markers and genetic testing (Figure ~\ref{fig2}). The adapted pedigree builder ensures that data entry is quick, efficient, and minimizes errors or omissions.

\subsection*{Fam3PRO Model}
We used Fam3PRO (version 2.0.1, previously released as PanelPRO) to estimate mutation carrier probabilities and future cancer risks \cite{Lee2021-xx}. The input of the model is a table of FHx where each row corresponds to an individual, and columns include sex, age, race, ethnicity, ancestry, individual ID, father and mother IDs, cancer history (as an indicator variable per cancer type), tumor markers, age of cancer diagnosis, surgical history (indicator per surgery), age of the surgery, and test results from gene panels. As such, the pedigree information is stored in a table format, and can be utilized by the Fam3PRO model effectively. During each function call, the code computes a likelihood matrix based on penetrances and allele frequencies relevant to the requested mutations and cancers, using a knowledge base on gene-cancer associations built from peer-reviewed studies. The model outputs are interactively visualized with the R package plotly (version 4.10.4) \cite{Sievert2020-lj}, and all results are available for download as PDFs. 

\subsection*{Data Safety and Confidential Information} To protect user credentials, we used the shinyauthr (1.0.0) \cite{Campbell2024-hm} and sodium (1.3.2) \cite{Ooms2024-pp} R packages inside the application to protect the user account. In our implementation sodium encrypts the password on the server so that the password is known only to the user. The RStudio Connect server where the Fam3PRO UI is deployed is hosted by Dana-Farber Cancer Institute. The UI website operates over the HTTPS protocol, ensuring that the connection between the user's browser and the server is encrypted, and users' data are protected. For security reasons, the interface does not provide any fields for entering sensitive or personally identifiable information (for example, name, zip code, and birth date), and users can permanently delete their saved pedigrees or genetic testing results for each relative at any time. Access to stored pedigree information is limited to the user, their designated manager (if applicable) and the system administrator, each of whom may retain a copy under their own account. The UI incorporates Google Analytics, which is only used to monitor the amount of available computing resources and the website traffic, without collecting other data. 

\section*{Results}

\subsection*{Pedigree Builder Results}

\begin{figure}[H]
  \centering
  \includegraphics[width=1.1\linewidth]{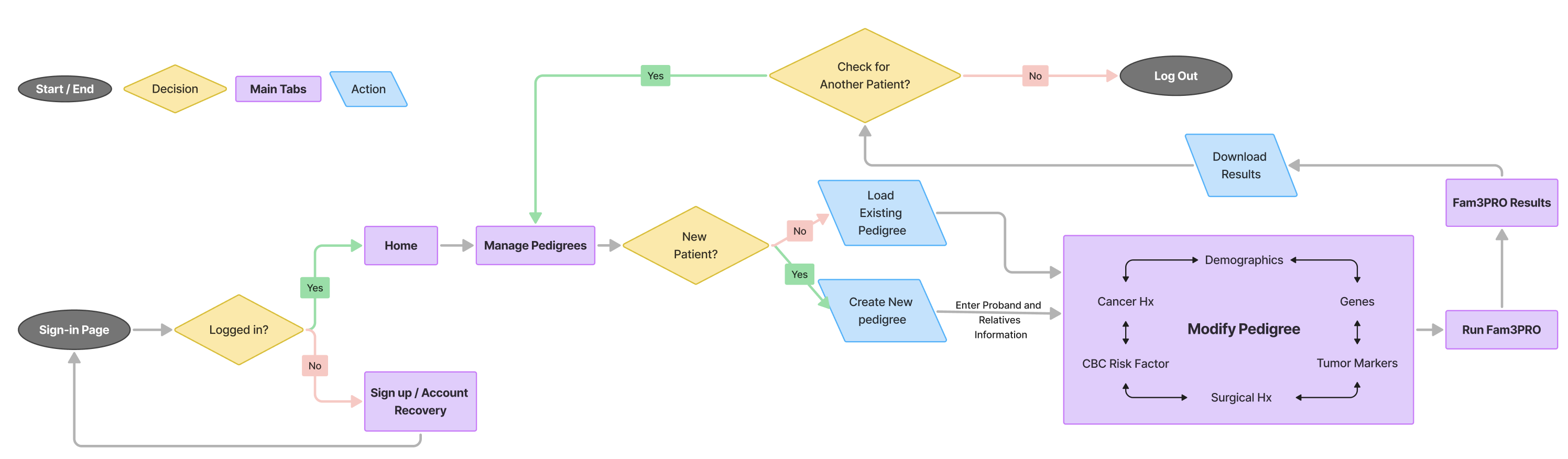}
  \caption{User journey of Fam3PRO UI. After creating an account, users can add and modify pedigree information, including additional risk modifiers and details. They can then run the Fam3PRO model to generate results. This workflow can be repeated for different pedigrees to assess multiple patients.}
  \label{fig2}
\end{figure}

To begin building a pedigree, a user first needs to designate a family member as the proband. The proband is the individual whose risk is being estimated based on their FHx and serves as the focal point. The user is required to enter the following minimum information: a unique pedigree ID, the proband's sex, and the proband's age. From here, the user completes the proband information, which is divided into tabs for demographic information, cancer history, prophylactic surgical history, and genetic testing (Figure ~\ref{fig2}, ~\ref{fig3}, ~\ref{fig4}). If the proband has had breast cancer in the past, there are additional tabs for assessing contralateral breast cancer (CBC) risk (Figure ~\ref{fig4}D) based on the cbcrisk R package (2.0) which has been incorporated into Fam3PRO \cite{Chowdhury2017-jq, Sajal2022-mg}. To estimate the risk of CBC, users can input information on the proband's first breast cancer (pure invasive, mixed invasive, or ductal carcinoma in situ, abbreviated as DCIS), their treatment history (including the use of anti-estrogen therapy), any history of high-risk pre-neoplasia, BI-RADS breast density results, and the size of the initial breast tumor. If the proband has had breast cancer or colorectal cancer, they will also be asked for the results of any associated tumor marker tests (Figure(~\ref{fig4}E)). Currently, users can input tumor testing results for breast cancer tumor markers, including ER, PR, HER2, CK5.6, and CK14. The demographic data collected, other than age and sex, include race and Hispanic ethnicity (which are used by Fam3PRO to adjust cancer penetrance values), Ashkenazi Jewish (AJ) and Italian ancestry (which are used by Fam3PRO to adjust BRCA1 and BRCA2 allele frequencies), and alive vs deceased status (Figure(~\ref{fig3}B)). The cancer history entries include the type of cancer and the age of diagnosis. Although Fam3PRO evaluates 17 different cancer types, the user can enter additional cancer types beyond the 17 to ensure a complete cancer history is stored. The surgical history inputs ask about any past bilateral mastectomies, hysterectomies, and bilateral oophorectomies along with the ages of those surgeries (Figure(~\ref{fig3}E)). The optional gene testing fields ask the user to first enter a single or multi-gene panel test which the user can select from a pre-established list or create their own custom panel, if the individual has undergone testing (Figure(~\ref{fig3}C)). Once the panel is created, the user can record the results of the genes in which variants were identified including pathogenic/likely pathogenic (P/LP) variants, variants of unknown significance (VUS), or benign/likely benign (B/LB) variants, along with the nucleotide, protein, and zygosity information for each (Figure(~\ref{fig3}D)). Any genes in the panel not marked as one of these three categories are assumed to be negative. Fam3PRO by default considers P/LP genes as causing cancer susceptibility, and B/LB genes are treated as negatives. VUS genes are collected to ensure a complete history, in the event they are upgraded to P/LP at a later date or downgraded.

Once the proband's information has been entered, the user may enter the number of proband's daughters/sons, sisters/brothers, maternal aunts/uncles, and paternal aunts/uncles to build the structure of the pedigree and family tree. Parents of the proband are added automatically, and if any aunts or uncles are indicated, the linking set of grandparents are also added automatically. After this step, the interactive pedigreejs family tree appears on the left side of the screen, while the R shiny data inputs are on the right side of the screen. Using the interactive tree, the user can easily add additional, more distant relatives to the pedigree such as cousins, great-grandparents, grandchildren, and nieces and nephews, along with any half-relations such as half-siblings. A drop-down menu allows the user to select any relative in the pedigree so that demographic information, cancer history, information of CBC risk based on primary breast cancer conditions and treatment, and tumor markers, surgical history, and genetic testing results can be entered, if available. It is recommended that the user enters at least the age, deceased status, and cancer history of each individual in the pedigree. Additional information will improve the model's performance but may be unknown to the proband such as a relative's age information. Fam3PRO uses a multiple imputation procedure with iterative sampling to estimate the age of individuals (if unknown) in the pedigree \cite{Lee2021-xx, Biswas2013-yv}. Every change in the pedigree is automatically saved to the database.

The user can also view the pedigree, including cancer history and genetic information, in table format if needed. The user also has the option to download the pedigree, which includes the full pedigree table, the image of the family tree, a detailed table of cancer history, a detailed genetic information table, and a data dictionary. At any time in the future, the user can return to their F3PI account, select a previously created pedigree and update it. They can also create copies of the pedigree under a different pedigree ID or permanently delete any of their pedigrees from the database.

\subsection*{Fam3PRO Input Results}

Once the user has created a new pedigree or loaded a previously created one, they can run the Fam3PRO model which has its own tab. The user can customize basic and advanced Fam3PRO settings under the tab "Run Fam3PRO" (Figure~\ref{fig5}). The basic settings include model selection, maximum number of genes with P/LP (maximum simultaneous mutations) per individual, and the year intervals for future cancer risk predictions (e.g. 1-year, 10-year, etc). The "peeling-paring" algorithm in Fam3PRO uses the maximum simultaneous mutations as its paring parameter for restricting the possible combinations of genotypes for each individual. This is set to a default value of 2, which provides an adequate approximation for clinical purposes as well as fast model performance \cite{Madsen2018-on, Lee2021-xx}. Note that the parameter could also be set to the number of distinct genes, which gives exact calculations. The model selection choices are Fam3PRO22 (which considers all 22 genes and 18 cancers currently in the model), Fam3PRO11 (a subset of 11 genes and 12 cancers), or the user can choose to specify a custom model involving any combination of the 22 genes and 17 cancers. Advanced settings allow customization of several options, including assignment of the race and ancestry category for all unknown races and ancestries, the number of iterations of age imputation for each missing age in the pedigree, whether future cancer risk should be estimated using crude (accounting for competing risks from other causes) or net 
(excluding deaths from causes other than the specified cancer) penetrance values, whether prophylactic interventions should be considered, whether to consider the proband's germline testing information in the calculations, whether to consider multiple variants in the BRCA1 and BRCA2 genes, and whether to allow Fam3PRO to automatically detect loops in the pedigree and break them by creating clone entries when necessary (Figure ~\ref{fig5}).

\subsection*{Fam3PRO Outputs Results}

Once the user runs the model, they are provided with the detailed information from Fam3PRO's built-in pedigree validation function (checkFam) on the "Console Output" tab, which provides information on age imputation and other assumptions (for example, germline testing results for certain genes are assumed to be heterozygous for any PV). The user is then presented with two interactive plots, one with mutation carrier probabilities and the other of future cancer risks (Figure ~\ref{fig6}A, ~\ref{fig6}C). The future cancer risk plot shows the estimated future cancer risk at the time intervals specified during model selections from the proband's current age to age 95. The plot also contains the cancer risks for a person at average risk from the SEER database 2022 \cite{National-Cancer-InstituteUnknown-ox}, which approximates the risk for non-carriers, in view of the infrequency of PVs in the general population. This allows the user to compare the proband's risk relative to the general population. The mutation carrier probabilities provide the probability of carrying a PV in each gene in the selected model, and their probability of being a non-carrier of any PV in the selected genes. As a reference, the graph includes a dashed red line at the probability threshold of 2.5\% which is a common cutoff used clinically to determine whether to recommend genetic testing to the proband. Each plot is accompanied by a data table with the plotted data (Figure ~\ref{fig6}B, \ref{fig6}D). The table also contains the joint probabilities that the proband is a carrier of multiple genes with PV if the selected model parameter allowed consideration of more than one PV per person. An additional table contains the model parameters used for traceability. The user can download all plots, tables, the pedigree in table format, and an image of the family tree in separate pdfs and one combined zip file with one click.

\begin{figure}[H]
  \centering \includegraphics[scale=0.8]{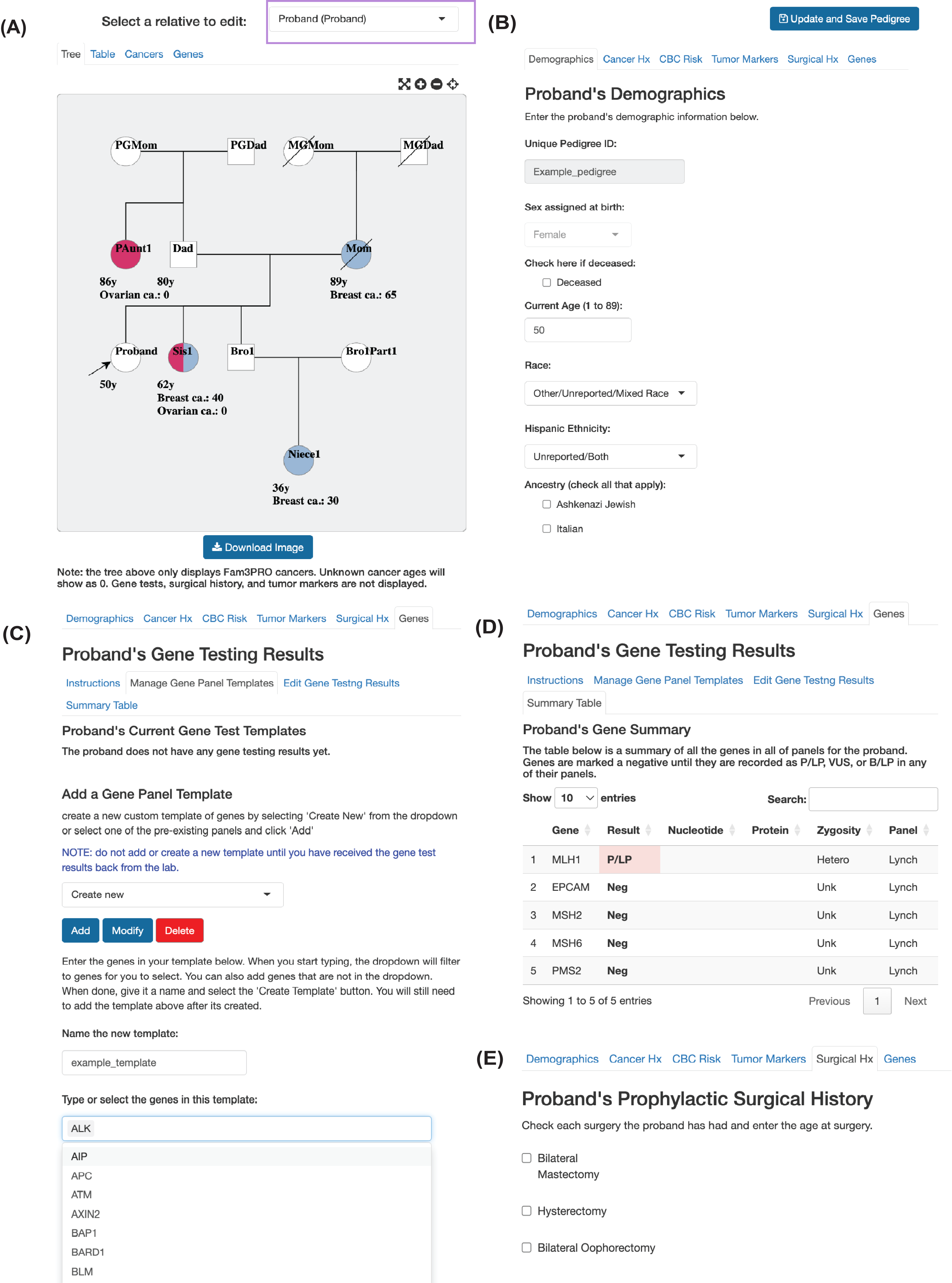}
  \caption{Example user input for proband-specific information: (A) Interactive pedigree tree using pedigreejs, currently editing the proband (highlighted in arrow and the rectangle box at the top), (B) demographics inputs for the proband, (C) page for the proband's germline mutation testing results, where users can create, load, and modify their own genetic testing panels, (D) summary table of the proband's gene testing results, and (E) prophylactic surgical history of the proband.}
  \label{fig3}
\end{figure}

\begin{figure}[H]
  \centering \includegraphics[scale=0.85]{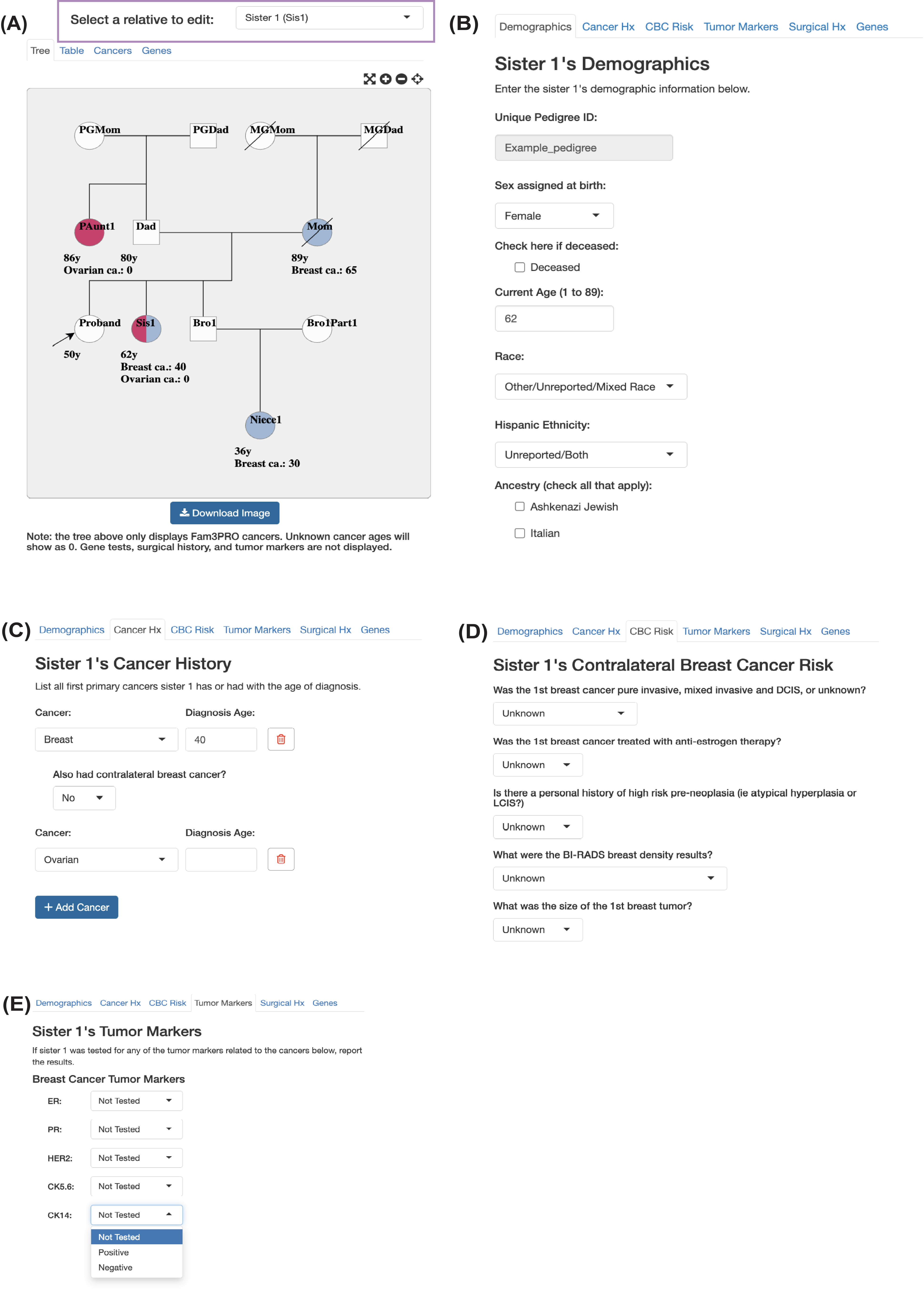}
  \caption{Example user input for relative-specific information: (A) Interactive pedigree tree using pedigreejs, currently editing proband's sister 1 (highlighted in arrow and the rectangle box at the top), (B) demographics inputs for the sister 1, (C) cancer history for sister 1, (D) contralateral breast cancer risk modifiers for sister 1, and (E) tumor markers for sister 1 (individual who has a cancer history}.
  \label{fig4}
\end{figure}

\begin{figure}[H]
  \centering \includegraphics[scale=0.62]{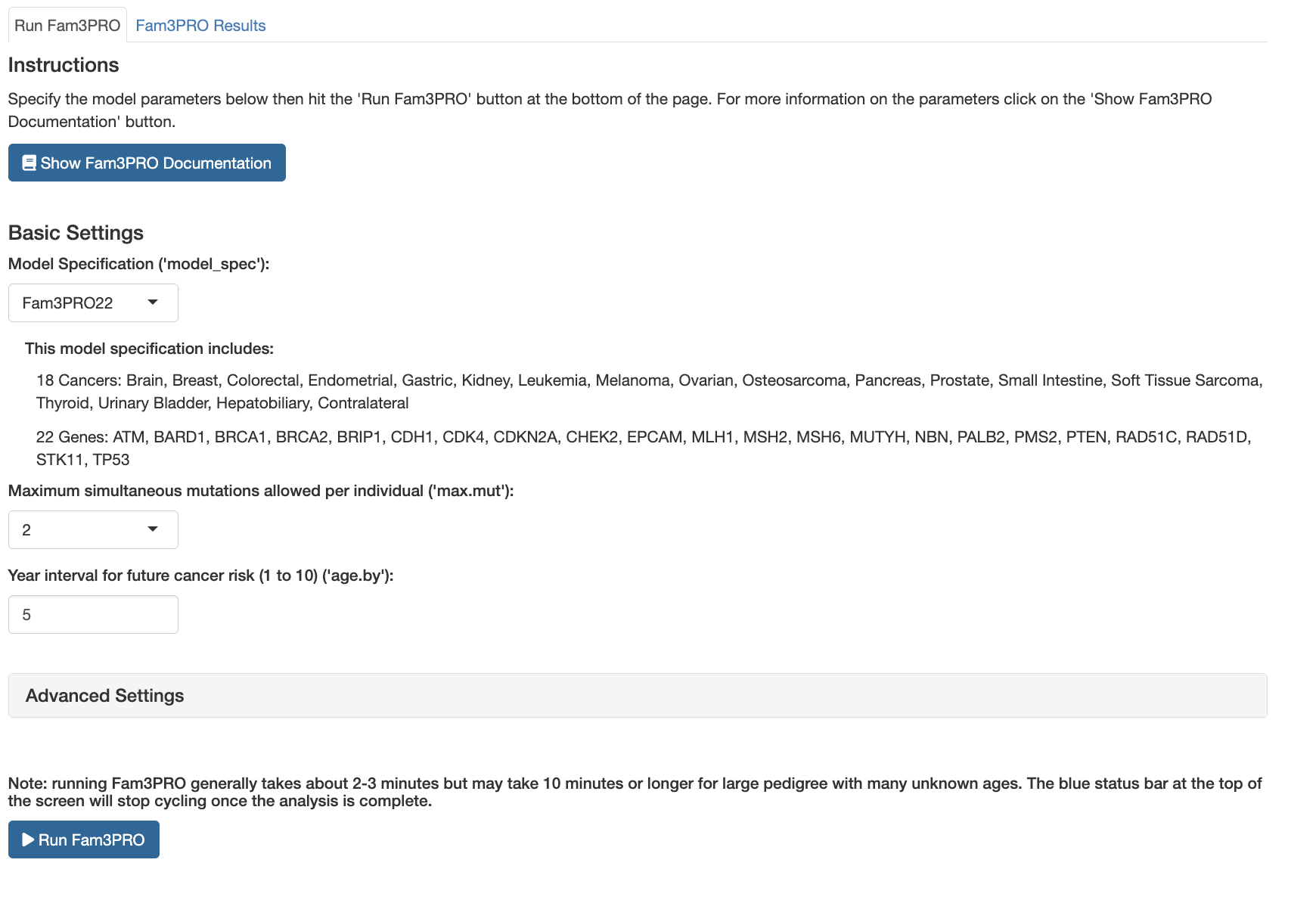}
  
  \caption{Fam3PRO model settings for running the example pedigree as shown in Figures ~\ref{fig3},~\ref{fig4}, including the specified model (default: Fam3PRO22) and both basic and advanced parameters.}
  \label{fig5}
\end{figure}

\begin{figure}[H]
  \centering
  \includegraphics[scale=0.88]{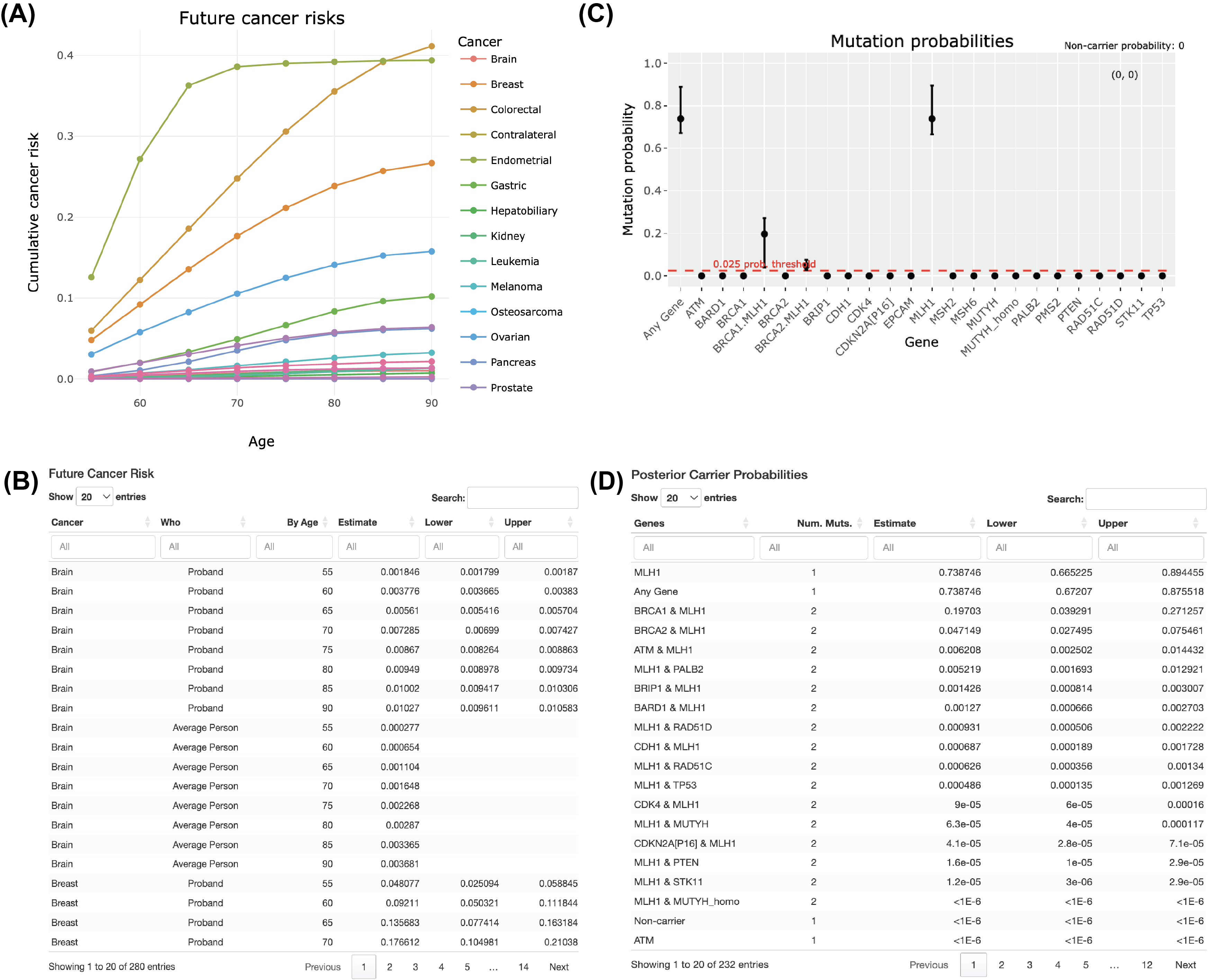}
  \caption{Fam3PRO predictions of future cancer risks (A, B) and carrier probabilities (C, D) for the proband in the same example pedigree as shown in the previous figures, in interactive figure and tabular formats.}
   \label{fig6}
\end{figure}

For most of the pedigrees, on average it takes 1 minute to run Fam3PRO. As family size increases, the number of relatives with missing age and/or number of iterations for age imputation increases, and more people use the tool simultaneously the compute time increases. Our Posit connect server has 12 CPU cores and maximum RAM of 30.9 GB. When being tested on the example pedigree as shown in Figures ~\ref{fig3} and ~\ref{fig4}, it takes around 28.6 seconds to run the model and display the results as shown in Supplemental Figure S1. Once the Fam3PRO computation begins, users will receive a message on the website providing an estimated completion time for their active session(s). An email notification will also be sent once the last model run for the same user is completed. The number of active sessions running the Fam3PRO model affects the total runtime until the final session completes, regardless of whether the sessions belong to the same user. The total waiting time increases linearly with the number of active sessions (Supplemental Figure S2). 

\section*{Discussion}

The F3PI is a publicly accessible tool for clinicians and genetic counselors to estimate a proband's future cancer risks for up to 18 cancers and mutation probabilities for up to 22 cancer-related genes using models from the Fam3PRO R package, without requiring prior programming experience. It enables clinicians to efficiently record detailed FHx in a format optimized for Fam3PRO, providing cancer risk and PV probabilities for multiple genes. The flexibility and comprehensiveness of F3PI has the potential to improve cancer prevention and early detection by identifying high-risk individuals and estimating both carrier status and future cancer risk.

Using pedigreejs, the interface provides user-friendly pedigree visualizations and allows interactive modifications for each relative in the pedigree \cite{Carver2017-uo}. Any changes are automatically saved to the database and reflected in the pedigreejs tree. Additionally, F3PI integrates comprehensive user input, allowing for detailed specification of each relative's information, including demographics, cancer and surgical history, tumor markers, and results of genetic testing, while also allowing model parameter adjustments to refine predictions.

F3PI could be used by clinicians with multiple patients, as well as by multiple collaborating clinicians within a clinical institution. The manager feature allows users with a manager role to access pedigree information from managed users, facilitating future data integration from multiple clinicians. Additionally, individual clinicians could use F3PI for research to conduct longitudinal studies on cancer risks, evaluating how well Fam3PRO predictions align with actual outcomes over time.

The main limitation of the application is the amount of time it takes Fam3PRO to run which is restricted by the compute resources of the server that hosts the website. There is also a slight delay, less than a second, when adding relatives using pedigreejs which is due to the time it takes pedigreejs to communicate with the R server to ensure the master pedigree which is stored in R, is synced. Additionally, F3PI does not support pedigrees with multiple probands when the focus of risk estimation is not for the proband but for other relatives.  

Future steps include speeding up the communication between pedigreejs and the R server, allowing the user to upload their own model inputs, such as cancer penetrance data in lieu of the native Fam3PRO penetrances. To facilitate usage, we may add customization options that allow clinicians to customize settings such their own genetic tests, assumed race, ethnicites, and ancestries to speed up data collection in areas where one characteristic is predominantly used (i.e., a clinical group in Italy could select a default Italian ancestry, unless the user specified otherwise). Each clinic could also be assigned its own database within MariaDB to ensure compliance with their institutional requirements. Future versions of F3PI are also expected to accommodate multiple probands within the same pedigree as supported in the Fam3PRO R package, providing cancer risk estimates and carrier probabilities for each selected individual in the pedigree.

\section*{Conflict of Interests}
Danielle Braun and Giovanni Parmigiani co-lead the Bayes-
Mendel Laboratory, which develops and maintains the BayesMendel
and PanelPRO software package. BayesMendel includes a variety of risk assessment
tools, including BRCAPRO, PancPRO, MelaPRO, MMRpro,
and PanelPRO/Fam3PRO, which are licensed for commercial use. None of the licensing fees for BayesMendel generate personal income for lab members. 

\section*{Acknowledgments}

The authors thank Dr. Stephen Gruber, Dr. Gregory Idos, and their team at City of Hope’s Center for Precision Medicine for their valuable feedback and suggestions during the development and testing of F3PI.

\section*{Author Contribution Statement}

GP and DB conceptualized and supervised the study.

XC drafted the manuscript, created figures, and contributed to F3PI development.

JK and LF contributed to F3PI development.

\section*{Code Availability}
The F3PI UI code is accessible through the github repository: \\
https://github.com/bayesmendel/Fam3PRO\_UI. 

\printbibliography

\end{document}


\maketitle

\clearpage 

\begin{figure}[H]
  \centering
  \includegraphics[scale=0.8]{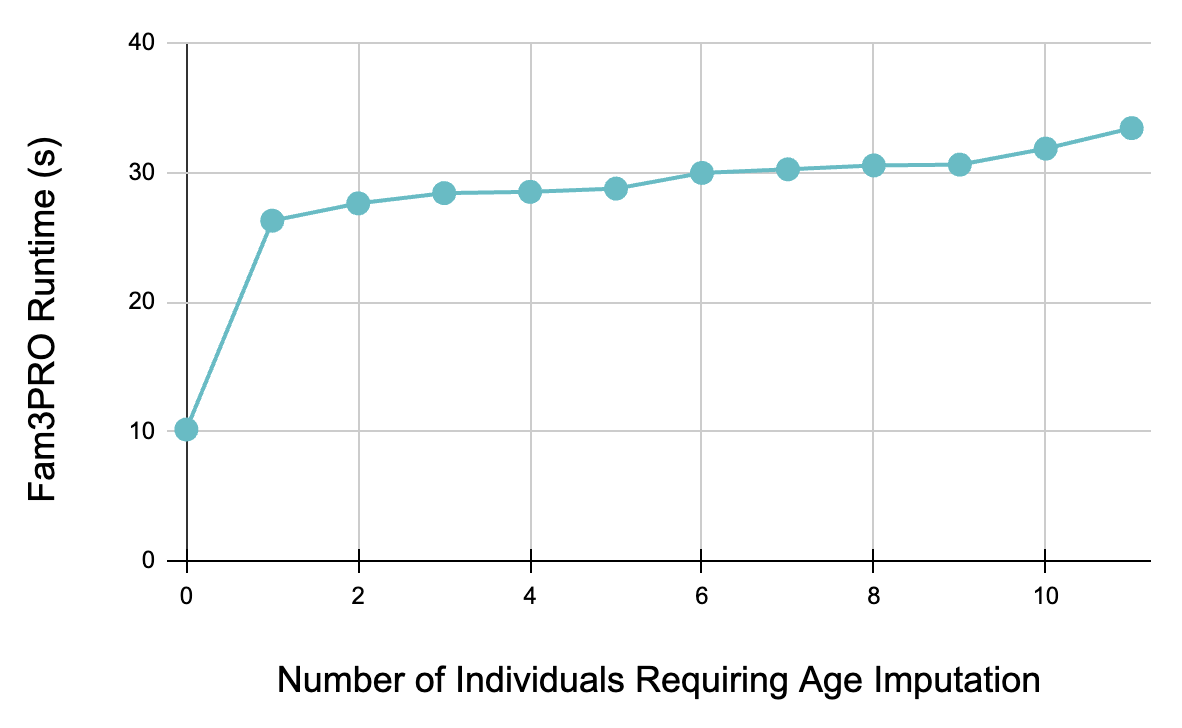}
  \caption{Fam3PRO runtime as a function of the number of relatives in the pedigree with missing ages (which require age imputations). The runtime experiments were conducted using F3PI on Posit Connect with an example pedigree consisting of 11 relatives in addition to the proband. The current age of the proband must be entered.}
  \label{suppfig1}
\end{figure}

\begin{figure}[H]
  \centering
  \includegraphics[scale=0.3]{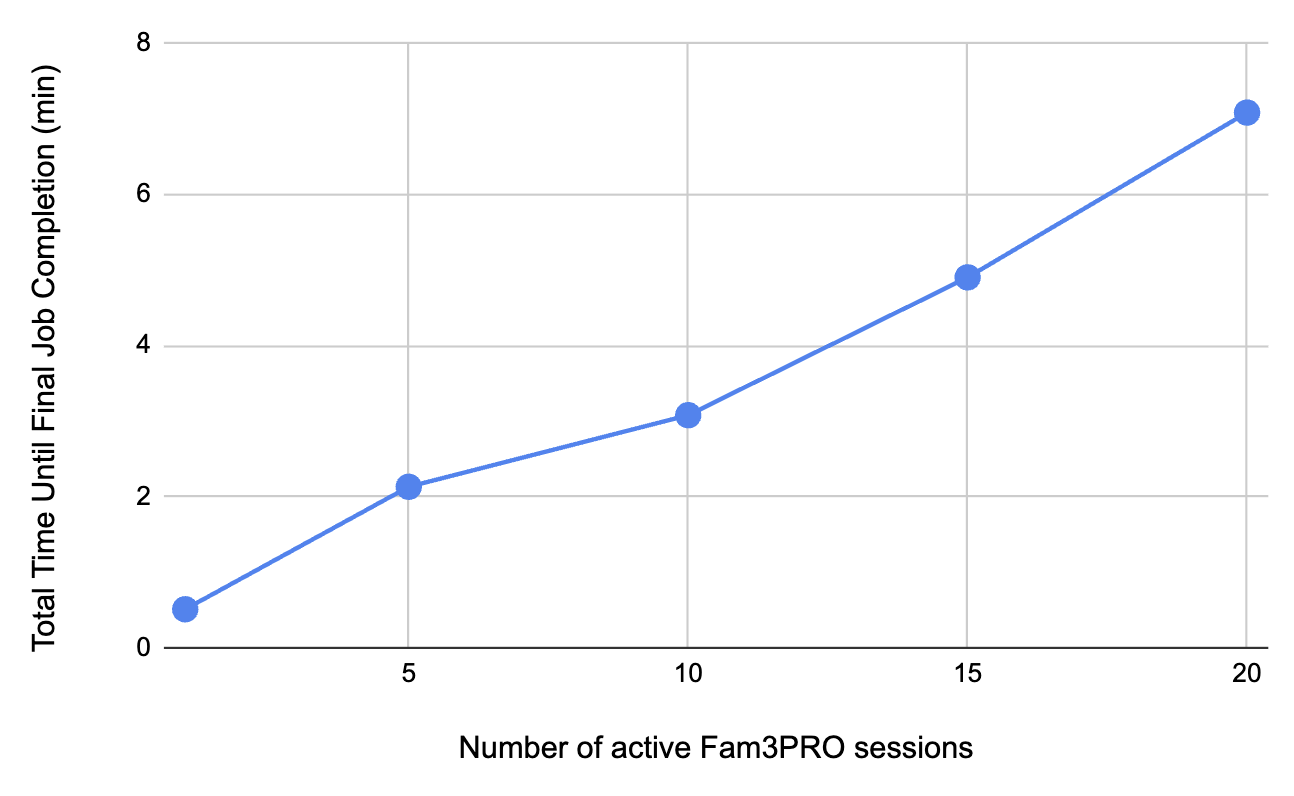}
  \caption{Fam3PRO runtime as a function of the number of concurrent sessions, using the same example pedigree as in previous figures but with all ages specified (no age imputations). Runtime is measured from the submission of the final job to its completion across all sessions.}
   \label{suppfig2}
\end{figure}